\documentclass[apj]{emulateapj}
\usepackage{graphicx}
\usepackage{txfonts}
\begin{document}

\title{21-cm synthesis observations of VIRGOHI 21 - a possible dark 
galaxy in the Virgo Cluster.}
\author{
Robert Minchin\altaffilmark{1} \altaffilmark{2}, 
Jonathan Davies\altaffilmark{2}, 
Michael Disney\altaffilmark{2},
Marco Grossi\altaffilmark{2} \altaffilmark{3},
Sabina Sabatini\altaffilmark{4}, 
Peter Boyce\altaffilmark{5}, 
Diego Garcia\altaffilmark{2} \altaffilmark{6},
Chris Impey\altaffilmark{7}, 
Christine Jordan\altaffilmark{8}, 
Robert Lang\altaffilmark{2},
Andrew Marble\altaffilmark{7}, 
Sarah Roberts\altaffilmark{2}, 
\and 
Wim van Driel\altaffilmark{9}}
\affil{}

\altaffiltext{1}{Arecibo Observatory, National Astronomy and Ionosphere Center,
Arecibo, PR 00612, US; rminchin@naic.edu}
\altaffiltext{2}{School of Physics and Astronomy, Cardiff University, Cardiff,
CF24 3YB, UK; Jonathan.Davies@astro.cf.ac.uk,
Mike.Disney@astro.cf.ac.uk, LangRH@cardifff.ac.uk,
Sarah.Roberts@astro.cf.ac.uk}
\altaffiltext{3}{Istituto di Fisica dello Spazio Interplanetario, via del Fosso
del Cavaliere 100, 00133, Rome, Italy; Marco.Grossi@roma1.infn.it}
\altaffiltext{4}{Osservatorio Astronomico di Roma, via Frascati 33, I-00040,
Monte Porzio, Italy; sabatini@mporzio.astro.it}
\altaffiltext{5}{Registry, Cardiff University, Cardiff, 
CF10 3UA, UK; BoyceP@cardiff.ac.uk}
\altaffiltext{6}{University of Bonn, Germany}
\altaffiltext{7}{Steward Observatory, University of Arizona, 933 N. Cherry 
Ave., Tucson, AZ 85721-0065, US}
\altaffiltext{8}{Jodrell Bank Observatory, University of Manchester, 
Macclesfield, Cheshire, SK11 9DL, UK; caj@jb.man.ac.uk}
\altaffiltext{9}{Observatoire de Paris, GEPI, CNRS UMR 8111 and Universit\'e 
Paris 7, 5 place Jules Janssen, F-92195 Meudon Cedex, France; 
Wim.vanDriel@obspm.fr}

\shorttitle{21-cm observations of VIRGOHI 21}
\shortauthors{Minchin et al.}

\begin{abstract}
Many observations indicate that dark matter dominates the extra-galactic 
Universe, yet no totally dark structure of galactic proportions has ever 
been convincingly identified.  Previously we have suggested that 
\object{VIRGOHI 21}, 
a 21-cm source we found in the Virgo Cluster using Jodrell Bank, was a 
possible dark galaxy because of its broad line-width ($\sim 200$ km\,s$^{-1}$) 
unaccompanied by any visible gravitational source to account for it.  We have 
now imaged VIRGOHI 21 in the neutral-hydrogen line and find what could be a 
dark, edge-on, spinning disk with the mass and diameter of a typical spiral 
galaxy.  Moreover, VIRGOHI 21 has unquestionably been involved in an 
interaction with \object{NGC 4254}, a luminous spiral with an odd one-armed 
morphology, but lacking the massive interactor normally linked with such a 
feature.  Numerical models of NGC 4254 call for a close interaction 
$\sim 10^8$ years ago with a perturber of $\sim 10^{11}$ solar masses.  This 
we take as additional evidence for the massive nature of VIRGOHI 21 as there 
does not appear to be any other viable candidate. We have also used the Hubble 
Space Telescope\footnote{Based on observations made with the NASA/ESA Hubble 
Space Telescope, obtained at the Space Telescope Science Institute, which is 
operated by the Association of Universities for Research in Astronomy, Inc., 
under NASA contract NAS 5-26555. These observations are associated with program
\#10562.} to search for stars associated with the H\,{\sc i} and find none 
down to an $I$ band surface brightness limit of $31.1 \pm 0.2$ mag 
arcsec$^{-2}$.
\end{abstract}


\section{Introduction}

The ability of a galaxy to form stars depends critically on the fraction of 
its mass that forms the baryonic gas disk and the temperature of this gas.  
High densities shield the gas from ionising radiation, allowing it to cool 
while higher intensity ionizing radiation keeps the gas hot.  Models
predict that galaxies can form with gas column densities that prohibit
star formation yet provide some self shielding from the ionizing
background (Davies et al. 2006). Such dark galaxies are potentially
detectable by blind 21-cm surveys of the sky. The possible existence of dark 
galaxies has previously been discussed by Jimenez et al. (1997), Hawkins 
(1997), Verde et al. (2002), and Davies et al. (2006); see also Taylor \& 
Webster (2005) for an alternative view.

Objects detected at 21-cm but with no optical counterparts have been
known about for many years; these include high velocity clouds
(Wakker \& van Woerden 1997), the Leo Ring (Schneider et al. 
1983), and various gas clouds
close to bright galaxies (Kilborn et al. 2000; Boyce et al 
2001; Ryder et al. 2001).  However,
none of these objects have the characteristics of a galaxy, i.e.
detectable emission over galaxy-sized spatial scales and a velocity
structure consistent with a rotating and gravitationally bound disk.  Recently
two candidate dark galaxies have been reported: VIRGOHI 21 (Minchin
et al. 2005) and HVC Complex H (Simon et al. 2006). A third possibility is the 
H\,{\sc i} cloud associated with the Local Group galaxy LGS3 (Robishaw et al. 
2002).

In this paper, we present high resolution H\,{\sc i} observations of VIRGOHI 
21, which we believe add further support to our hypothesis that it is 
a dark galaxy.  We also describe HST observations that were specifically 
designed to search for red giant stars at the distance of the Virgo Cluster 
and so place very faint limits on the surface density of stars that might be 
associated with VIRGOHI 21.

\section{21-cm Observations and Analysis}

The new 21-cm data were taken in March 2005 at the Westerbork Synthesis Radio 
Telescope (WSRT) in two full 12-hour syntheses and reduced using the
{\sc Miriad} package.  The data were flagged for shadowing and on two of the 
fourteen 25-meter antennae one polarization was flagged due 
to problems with the gain.  A spectral bandwidth of 10 MHz covered the velocity
range 930 - 3070 km\,s$^{-1}$.  Removal of the noisier end channels left 230 
useful channels of width 8.2 km\,s$^{-1}$ each, giving a velocity resolution 
of 10 km\,s$^{-1}$ over the range 980 to 2890 km\,s$^{-1}$.  Continuum removal 
was carried out in the UV plane using {\sc uvlin}.  The standard source 3C147
was used for calibration.  Cleaning used a robust setting of 1, close to 
normal weighting (Briggs 1995). The cleaned cube was Gaussian 
smoothed spatially and Hanning smoothed in velocity.  This was then used as
a template for regions where flux was present in the data in order to carry 
out a deeper cleaning of the dirty image, which gave the final cube used in 
the analysis.  The synthesized beam was $99\arcsec\times 30\arcsec$ in size 
(extended north-south) and the noise was 0.3 mJy\,beam$^{-1}$\,channel$^{-1}$, 
giving a $5\sigma$ column-density limit for sources 25 km\,s$^{-1}$ wide of 
$2\times 10^{19}$ atoms cm$^{-2}$.
%\clearpage
\begin{figure*}
\plotone{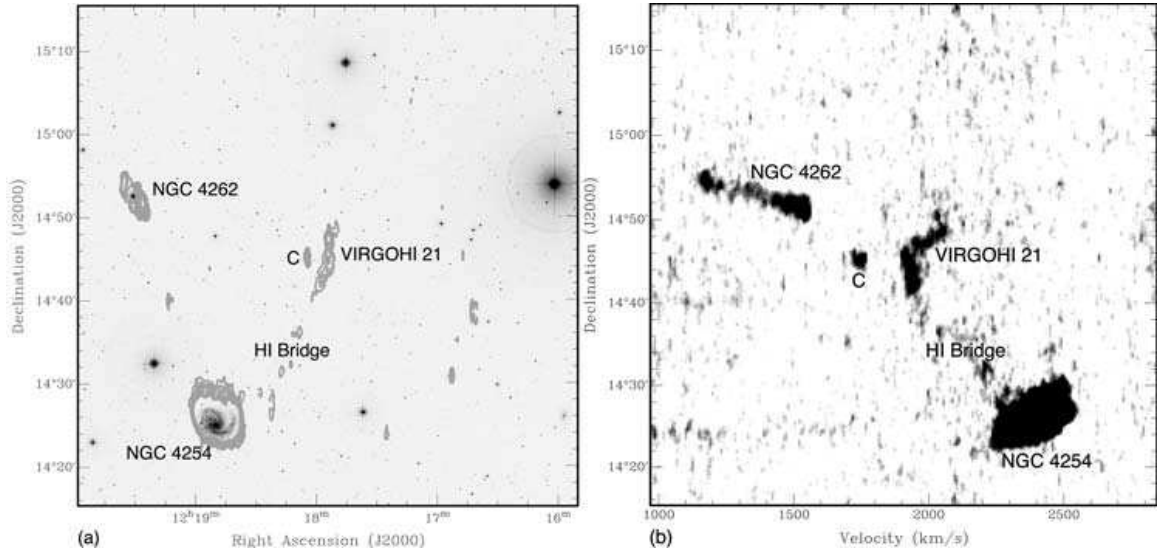}
%\resizebox{\columnwidth}{!}{\includegraphics{f1.eps}}
\caption{(a) H\,{\sc i} contour map of the 21-cm observations, superimposed
on a 1 square degree negative Digitized Sky Survey image. Contours are from
$2.5 \times 10^{19}$ to $2 \times 10^{20}$ cm$^{-2}$ at intervals of $2.5 
\times 10^{19}$ cm$^{-2}$.(b) Shows the declination-velocity projection of 
the data cube.  More detail can be seen in the online animation of the 
datacube (Fig. \ref{animation}).}
\label{fig1}
\end{figure*}
%\clearpage
Fig. \ref{fig1}a shows a neutral hydrogen (H\,{\sc i}) contour map of the 
field superimposed on a negative optical image.  VIRGOHI 21 is the elongated 
structure in the center (which is at about 2000 km\,s$^{-1}$).  A faint bridge 
can be seen stretching down to the prominent spiral NGC 4254 
(2400 km\,s$^{-1}$) while the other two sources, \object{NGC 4262} 
(1350 km\,s$^{-1}$, 
upper left) and the faint galaxy `C' (1750 km\,s$^{-1}$, immediately to the 
left of VIRGOHI 21) appear unconnected.  Fig. \ref{fig1}b shows the 
velocity-declination
projection of the full 3 dimensional data cube.  Now NGC 4254 is at the bottom 
right while VIRGOHI 21 is the angular structure in the center with `C' to its 
left.  Far more detail can be seen in the animation of the data cube, Fig. 
\ref{animation} (available in the electronic edition of the {\it Astrophysical 
Journal}, only the first frame is shown in the print edition).  The 
bridge between VIRGOHI 21 and NGC 4254 is clear, as is the lack of any 
connection between VIRGOHI 21 and either NGC 4262 or galaxy `C'.  The apparent 
alignment of NGC 4262 with VIRGOHI 21 and NGC 4254 in Fig. \ref{fig1}b is 
merely a consequence of the particular projection shown.
%\clearpage
\begin{figure}
\plotone{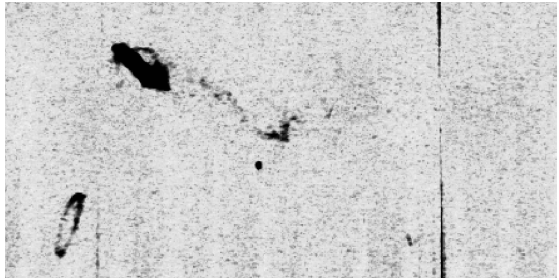}
\caption{This figure is available as a color mpeg animation in the electronic 
edition of the {\it Astrophysical Journal}, only the first frame is shown 
here.
The figure shows an animation of the WSRT data cube,
oriented with increasing velocity upwards and
rotating around this axis.  VIRGOHI 21 can be seen in the center joined by 
a stream of neutral hydrogen to NGC 4254 (above and to the side).  Object 
C is below VIRGOHI 21 and NGC 4262 is the ring structure near the bottom.  
The vertical column is the continuum source 1214+1456 (White \& 
Becker 1992) which was incompletely removed by {\sc uvlin}.  The
faint vertical striping is due to incompletely removed continuum from 
M87, which is detected in the sidelobes of the array.}
\label{animation}
\end{figure}
%\clearpage
The bridge is $\sim$ 25 arcmin long and stretches from 2250 km\,s$^{-1}$
at $+14^\circ28^\prime$ at the low-velocity (western) edge of NGC 4254
to 1900 km\,s$^{-1}$ at $+14^\circ 41^\prime$.  Here it meets a section 
of relatively high column-density H\,{\sc i} which is flat in velocity until 
$+14^\circ 46^\prime$ where there is a strong velocity gradient in the
opposite direction from that seen in the bridge, rising to 2100 km\,s$^{-1}$ 
at $+14^\circ 49^\prime$.

\begin{figure*}
\centering
\plottwo{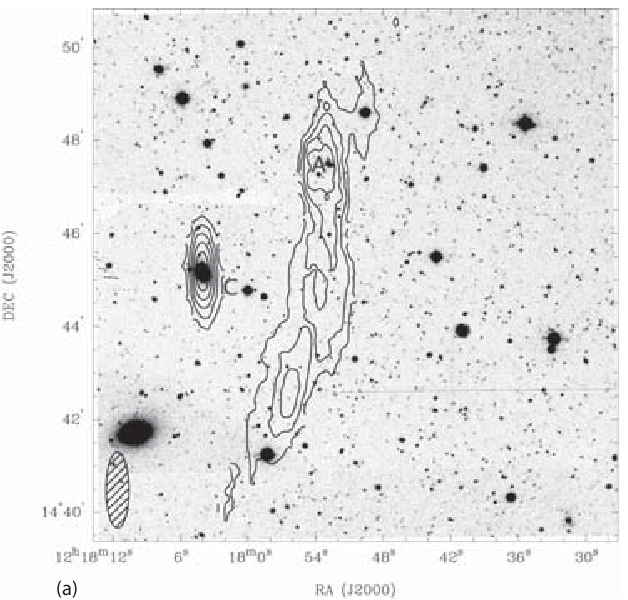}{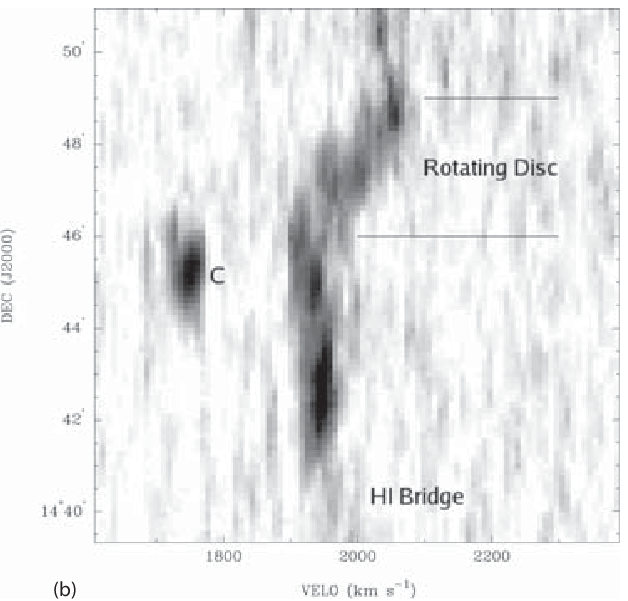}
\caption{As Fig. \ref{fig1}, but expanded and superimposed on a negative of 
our deep
CCD $B$-band image (Minchin et al. 2005) with a surface-brightness 
limit of 27.5 $B$ mag. arcsec$^{-2}$.  The hatched ellipse in the bottom-left 
corner indicates the WSRT half-power beam.}
\label{fig2}
\end{figure*}

Fig. \ref{fig2} is a blow-up of the source region superimposed on a far 
deeper CCD optical image.  The CCD image was made by combining two 
$B$-band images: a 750 s Isaac Newton Telescope (INT) Wide Field Camera (WFC)
survey image taken during photometric conditions in March 2000 
and our own 600 s image taken with the same instrument during non-photometric 
conditions in May 2004. Both sets of data were reduced using the Wide Field 
Survey 
pipeline\footnote{http://www.ast.cam.ac.uk/$\sim$wfcsur/technical/pipeline/} 
which includes 
de-biasing, bad pixel replacement, non-linearity correction, flat-fielding 
and gain correction.  The images were combined using the Starlink task 
{\sc MakeMos} in the {\sc CCDPack} package and photometry was taken from the 
WFC survey image, which has a photometric accuracy of 0.05 mag.  Combined,
the two images have a $1\sigma$ sky noise of 26.7 mag arcsec$^{-2}$ which has 
been improved to 27.5 mag arcsec$^{-2}$ by binning into 1 arcsec pixels. This 
gives a surface brightness detection limit of about 27.5 mag arcsec$^{-2}$ 
for an object of diameter 10 arcsec.  We find no optical counterpart down
to this surface brightness and size limit.

\begin{figure}
\centering
\plotone{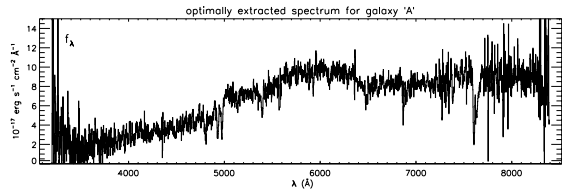}
\caption{Optical spectrum of object `A' taken with the 6.5-m MMT.  The H and 
K lines give a redshift of $z=0.2535$.}
\label{fig3}
\end{figure}

There is a small, faint galaxy labeled `A' superposed on the highest
H\,{\sc i} contour ($10^{20}$ cm$^{-2}$) at declination 
$+14^\circ47.4^\prime$. An optical spectrum from the 6.5-m MMT in Arizona 
shows that it is at a redshift of $z=0.25$ and is therefore unconnected 
with VIRGOHI 21 (see Fig. \ref{fig3}).  The 17th magnitude galaxy `C' to the 
left at $+14^\circ45^\prime$ is an H\,{\sc i} point-source at this resolution. 
By comparison, VIRGOHI 21 is an extended structure in both dimensions rather 
than a collection of discrete compact clouds.  The velocity-declination plot 
(right) shows the complex kinematic structure of VIRGOHI 21.  The most 
remarkable feature is the tilted portion between $+14^\circ46^\prime$ 
(1900 km\,s$^{-1}$) and $+14^\circ49^\prime$ (2100 km\,s$^{-1}$) which 
resembles the signature of an edge-on rotating disk (e.g. Kregel et al. 2004).
A tentative detection of gas further to the north (in Fig. 3b) could either be 
part of the interaction or clumps further out in the disk. Note the lack of 
connection between galaxy `C' and VIRGOHI 21.
%\clearpage

\begin{figure*}
\centering
\epsscale{.77}
\plotone{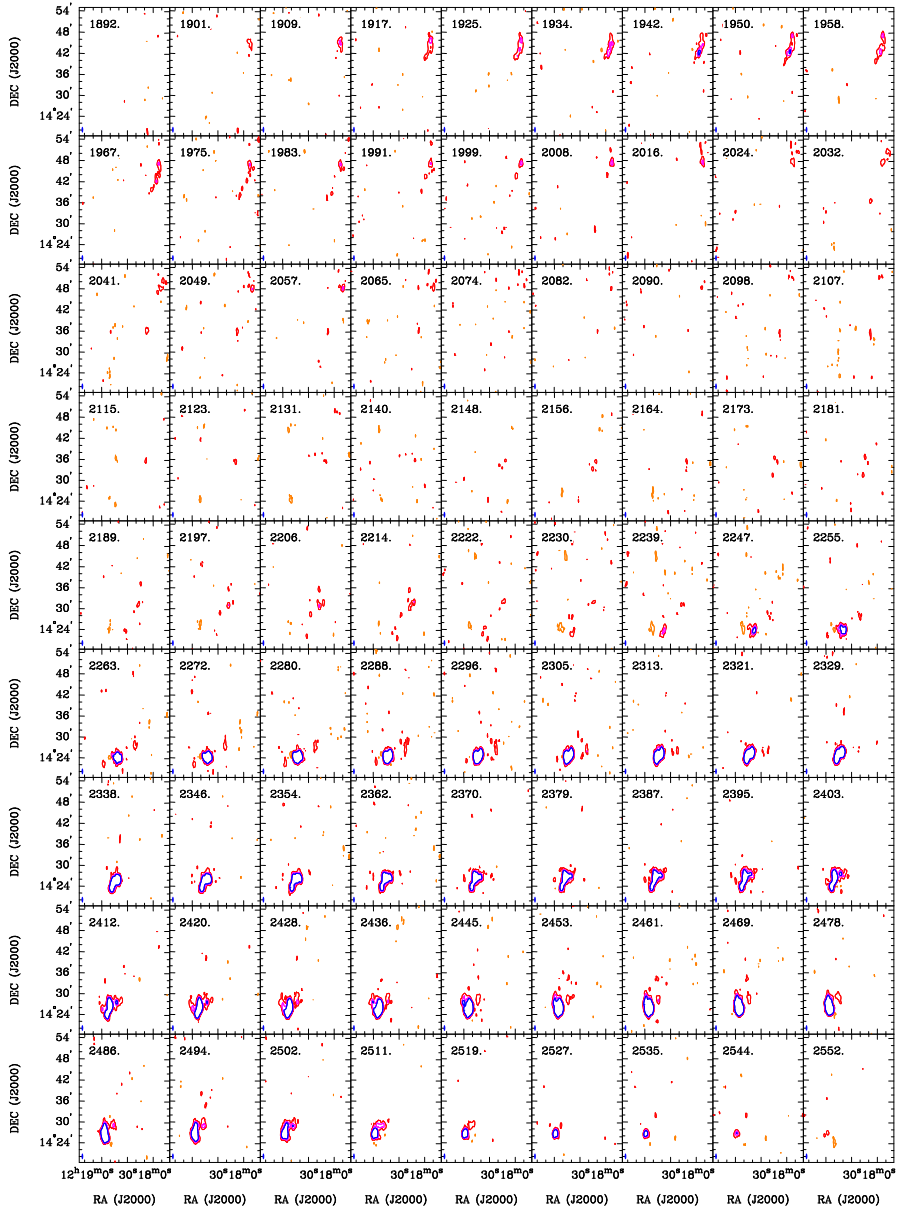}
\caption{Channel maps of the VIRGOHI 21 -- NGC 4254 system.  Contours are at
$-1$ mJy ($-3\sigma$), 1 mJy ($3\sigma$), 2 mJy ($6\sigma$) and 3 mJy 
($9\sigma$) (colored orange, red, magenta and blue in the electronic 
edition).  The beam size is shown by the ellipse in the bottom-left and
the velocity of each channel (in km\,s$^{-1}$) is given in the top-left.
See the electronic edition of the Journal for a color version of this figure.}
\label{fig4}
\end{figure*}

\begin{figure*}
\centering
\rotatebox{270}{\includegraphics[width=0.85\linewidth]{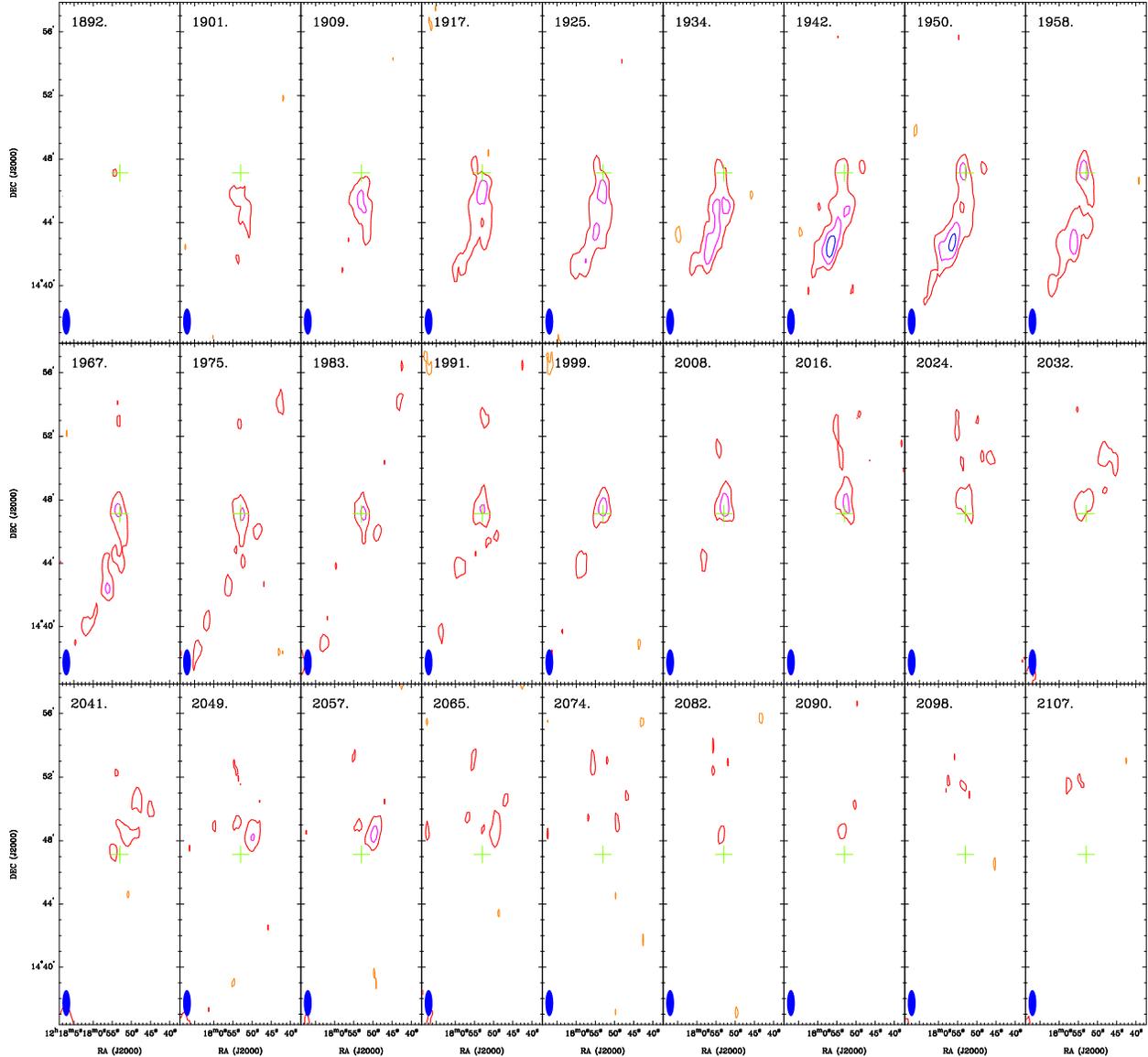}}
%\plotone{f6_color.eps}
\caption{Channel maps of the central region of VIRGOHI 21, contours as in Fig.
\ref{fig4}. The cross indicates the peak of the zeroth moment map, which is at 
the center of the rotating component (see Fig. \ref{fig2}).  It can clearly be 
seen that this component (declination range $14^\circ46^\prime$ to 
$14^\circ49^\prime$) starts south of this point and moves northwards across it
with increasing velocity (as might be expected for rotation).  See the 
electronic edition of the Journal for a color version of this figure.}
\label{fig5}
\end{figure*}
%\clearpage
Figs. \ref{fig4} and \ref{fig5} show the channel maps from the WSRT 
observations.  Fig. \ref{fig4} shows
the whole VIRGOHI 21 -- NGC 4254 system, including the bridge connecting
the two which is only a $3\sigma$ detection in the individual channels but,
taken over all channels, is definitely present (see Fig. \ref{fig1}b and the 
animation of the datacube, Fig. \ref{animation}).  Fig. \ref{fig5} shows 
the cube between RA $12^h17^m46^s$ and 
$12^h18^m03^s$ over the velocity range of VIRGOHI 21.  It can be seen that the
section identified in Fig. \ref{fig2}b as the `Rotating Disc' undergoes a 
smooth shift northwards with increasing velocity, which resembles (taking into 
account
the difference in beam sizes) that observed in edge-on rotating systems 
(e.g. Uson \& Matthews 2003; Rupen 1991). Note that this section of VIRGOHI 21
is unresolved in individual channels, leading to a broadening of the linewidths
down each line of sight to $\sim 100$ km\,s$^{-1}$ and a lowering of the
column-density as the H\,{\sc i} is smeared by the beam.

\section{HST Observations}

To place a better limit on the presence of stars associated with VIRGOHI 21 we 
have obtained HST data
with a total exposure time of 23142 s.  The ACS Wide Field Camera was used 
with the $I$-band F814W filter. The pointings were centered on the H\,{\sc
i}-flux-weighted position of $12^h17^m53^s, 
+14^\circ45^\prime25^{\prime\prime}$ (J2000), 
with the ACS field of view matching the minimum H {\sc i} size (3.5
arcmin) very closely. Each orbital exposure was {\sc cr-split} in two for 
cosmic ray removal and was slightly offset to permit drizzling.

Preliminary data reduction was done by the Space Telescope Science
Institute with the standard pipeline. The flat-fielded
images have been combined with the {\sc Iraf} task {\sc Multidrizzle} to
obtain a distortion-corrected and drizzled final image. The pixel
size in the resulting images is 0.05 arcsec.

With this depth and spatial resolution we should be able to
resolve individual stars in VIRGOHI 21 down to $I \simeq 28$ and thus to
detect the tip of the Red Giant Branch which at the distance of
the Virgo Cluster would be close to 27 in $I$. The presence of
individual stars would not  conflict with the ground-based INT
observations: the $B$-band surface-brightness limit of 27.5 mag
would (for $B - I \simeq 2$) equate to around 0.5
detectable RGB stars per square arcsecond, or 2500 - 3000 over the
area of VIRGOHI 21.

Stellar photometry was performed with the {\sc Daophot} package in {\sc Iraf}
(Stetson 1987). A preliminary selection of stars with a threshold
of 3.5 sigma was performed with the automatic star-finding
algorithm {\sc Daofind} and their magnitudes obtained with the task {\sc Phot}
using an aperture radius of 3 pixels. A spatially variable model
point-spread function (PSF)  -- with a full width at half maximum
of 2 pixels --  was built with the task {\sc Psf} by selecting several
isolated and bright stars. Finally, {\sc Allstar} was used to fit the
model PSF to the stars in the input list and to produce the final
catalog.  The conversion to Johnson {\it I} magnitudes and the
aperture correction were done following Sirianni et al. (2005).  
Artificial star tests were used to determine the completeness of our
selection; the 50\% completeness limit is found to be $I=27.9$ mag.

\begin{figure}
\plotone{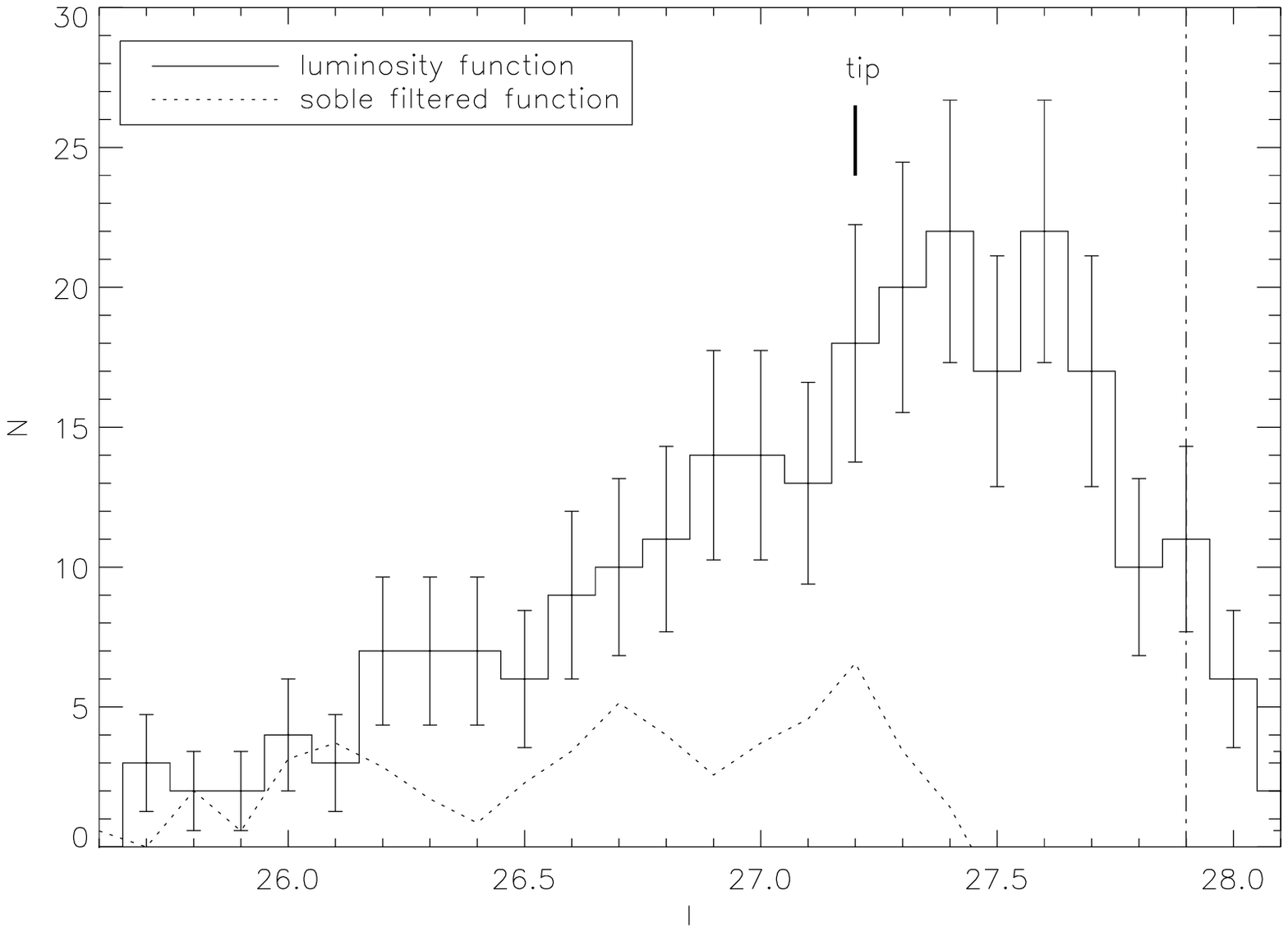}
\caption{The luminosity function of the final
list of stars. The dotted line shows the output of the edge
detection filter used to detect the tip of the RGB. The maximum of
the function indicates the position of the tip. The vertical line
at $I = 27.9$ mag indicates the 50\% completeness limit. }
\label{lum_func}
\end{figure}

We found that the photometry catalog produced with
{\sc Daophot/Allstar} ($\sim$600 objects) included  bad measurements due to
misidentifications such as  parts of bright background galaxies or
unresolved background galaxies whose contribution to the
luminosity function is significant at fainter magnitudes (see
Durrell et al. 2002). Given the long exposures required by such
observations we did not ask for additional time to look at another
field to determine the background density of star-like objects;
instead, we have removed objects from our list by
individually examining their radial profiles and rejecting those with a
FWHM $\gtrsim$ 2.5 - 3 pixels, corresponding to the value of the
PSF built with the brightest stars in the field.  The final list
consists of 281 stars and the corresponding luminosity function is
shown in Fig. \ref{lum_func}.
%\clearpage
%\clearpage
To estimate the contamination due to foreground stars, we compared
the number of detections  with a model of the expected number of
Milky Way stars in the position of the Virgo Cluster (Castellani 
et al. 2002; Cignoni et al. 2003).  This model shows
that the contribution of foreground stars is only significant at
brighter magnitudes ($I \leq 26.5$) where the number of stars
detected is small (Fig. \ref{lum_func}); at fainter magnitudes the larger
number of detections makes the foreground unimportant.

To search for the Tip of the Red Giant Branch (TRGB) we have applied a
Sobel filter to the binned luminosity functions.  The luminosity
function shows a slope change at $I = 27.2\pm 0.1$ mag, 
which we identify as the possible tip of the red giant branch.  This would 
place these stars at a distance of 16.6 - 18.2 Mpc, which is consistent with 
them being in the Virgo Cluster (d=16 Mpc), though the statistics are poor due 
to the relatively low numbers of stars in each magnitude bin. If  we assume 
this TRGB to be real, despite the low signal to noise ratio (around 2$\sigma$
on the Sobel-filter output), we can set the most stringent limit on the 
surface-brightness of VIRGOHI 21.

Assuming (very generously) that all of the fainter stars are also at the 
distance of the cluster 
(119 with $I = 27.2 - 27.9$) they have a combined magnitude
of $I = 22.3$, giving a surface brightness of 33.8 $I$
mag arcsec$^{-2}$. Using the formula of Durrell et al. (2002) gives a
correction for only having the top 0.7 mags of the RGB of $1.5 \pm 0.1$
mag arcsec$^{-2}$.  We do not detect any Asymptotic Giant
Branch stars; based on the work of Durrell et al. these could contribute
another $0.2 \pm 0.1$ mag arcsec$^{-2}$.  Overall, this gives
an upper limit for the total $I$-band surface brightness of 
$31.1 \pm$ 0.2 mag arcsec$^{-2}$. For normal galaxy
colors of $B-I = 1.5 - 2$, this equates to an extremely faint $B$-band surface
brightness of 32.4 -- 33.3 mag arcsec$^{-2}$.  

These results can be compared with studies of Virgo Cluster inter-galactic 
stars (Durrell et al. 2002) and those used to determine the 
color-magnitude diagrams of
dwarf galaxies belonging to the cluster (Caldwell 2006). Durrell et al.
imaged a field at about 40$^\prime$ northwest of M 87 ($\approx$190 kpc)
while the two fields observed by Caldwell were also relatively close to the
center of Virgo, one at $1.4^{\circ}$ north ($\approx$400 kpc)
and  the other at $1.1^{\circ}$ west ($\approx$315 kpc). These fields are much 
closer to M 87 at the center of the cluster
than VIRGOHI 21 and the number density of stars found is also much higher 
however, there is an apparent decrease in the number of stars 
as one moves away from the cluster center (see Fig. 9 in Caldwell 2006 and Fig.
6 in Durrell et al. 2002).  A comparison between the estimated
surface brightnesses of the three fields shows that $\mu_I$ ranges
from 27.7 $I$ mag arcsec$^{-2}$ (Durrell et al. 2002) for the
fields closer to the cluster center to $\mu_I = $ 27.9 and 29.9 $I$ mag 
arcsec$^{-2}$ (Caldwell 2006) for the more distant ones.  Our $I$ band
surface brightness of 31.3, at a projected distance of 1.1 Mpc from the
cluster center (M87), is thus consistent with the known population 
intra-cluster stars in
Virgo and so there is no evidence to suggest that our detections are 
associated with VIRGOHI 21 in any way.

\section{Discussion}

In the previous sections we have described our new observations of VIRGOHI 21. 
In this section we try and put them in context and particularly we want to 
assess the evidence for and against the dark galaxy hypothesis.
\begin{enumerate}
\item If attributed to gravitation, changes in velocity of galactic size over
galactic scales, as seen here, require masses of galactic proportions.  On
dimensional grounds the abrupt change in velocity $\Delta V$ ($\sim 200$ 
km\,s$^{-1}$)
seen in VIRGOHI 21 over a conservative length-scale $\Delta x$ ($\sim 14$ kpc 
$\simeq 5\times 10^{22}$ cm $\simeq 3$ arcmin at the Virgo Cluster distance
of 16 Mpc)
implies a mass $M \geq (\Delta V)^2 \Delta x/G \simeq 10^{10-11}$ solar masses 
if this is due to an object rotating in dynamical equilibrium. If the 
velocities are not due to rotation then changes in velocity like this can be 
observed if, for example, we are observing an arc of gas and our view is almost
 edge-on. These geometrical effects can lead to large velocity widths that may 
resemble rotation (e.g. Bournaud et al. 2004). The problem with this latter 
scenario (as also pointed out by Vollmer et al. (2005) is that the perturbing 
galaxy should be moving in the plane of the arc and so should be projected on 
the sky very close to it. We can simply quantify what we mean by this. Imagine 
two galaxies with radial velocities, $V_1$ and $V_2$ at either end
of an approximately linear tidal bridge of physical length $d$ pitched at an 
angle $\theta$ to the line of sight.  A telescope pointed towards it has a 
transverse beam diameter of $b$ at the bridge.  The only significant
gas motions within the bridge will be streaming velocities along its length.
From end to end of the bridge the radial velocity difference is $|V_2 - V_1|$
while within the telescope beam the measured velocity width, $\Delta V$, will 
be ($b/d\sin\theta)\times|V_2 - V_1|$.  But, bridges of any appreciable
size arise only when the total velocity difference between the 
interacting galaxies i.e. $|V_2 - V_1|/\cos\theta$, is of order 
the circular velocity $V_c$ in the galaxies involved (Toomre \& Toomre 1972).
It follows that $\Delta V/V_c \simeq (b/d)\cot\theta$.  Thus broad
line widths $\Delta V \simeq V_c$, as here, can only be seen within a beam if
both interactors appear to lie within, or very close to the beam
i.e. $d\sin\theta \leq b\cos\theta \leq b$.  We can find no potential 
perturber within the 
Arecibo beam ($b\simeq 3\farcm6$) or close to it. Thus we do not believe
VIRGOHI 21 can be tidal debris from an interaction of this nature. 

\item The existence of the H\,{\sc i} bridge indicates that NGC 4254 has 
undergone some kind of interaction. 
NGC 4254 is a luminous one-armed spiral galaxy 
sufficiently peculiar to have attracted several studies 
(Iye et al. 1982; Phookun et al. 1993; Vollmer et al. 
2005). According to Vollmer et al. 2005) 
the morphological peculiarities of NGC 4254 can be explained by an
interaction with a perturbing mass of $\sim 10^{11} M_\odot$.  As it is now 
clear
that VIRGOHI 21 is involved in this interaction, it should be considered a 
candidate for the perturber, particularly as no other candidate can be
easily identified.    The 
Vollmer et al. simulation implies an 
interaction $3\times 10^8$ years ago.  As the projected length of the bridge
is 120 kpc this would imply it has been drawn out at a projected speed of 390 
km\,s$^{-1}$, which is comparable to the radial velocity difference 
between NGC 4254 and VIRGOHI 21 of 400 km\,s$^{-1}$. Thus if VIRGOHI 21 were 
massive enough ($\sim 10^{11}$ $M_{\odot}$), it could be the cause of the 
tidal bridge.

\item If object C, the H\,{\sc i} galaxy just to the east (left) of VIRGOHI 21 
(Fig. \ref{fig2}) were involved its mass (see (a) above) should be 
$\sim 10^{11} 
M_\odot$.  However, 
its measured H\,{\sc i} velocity distribution and size suggest a probable mass 
$\leq 10^9 M_\odot$, while it has a luminosity of only $10^8 L_\odot$ and a 
$M_{HI}$ of $2\times 10^7 M_\odot$.  Galaxy `C' appears to be two orders of 
magnitude too 
undermassive and underluminous to explain the one-armed spiral structure of 
NGC 4254 and the H\,{\sc i} bridge connecting to it. 
 So, it seems very unlikely that `C' has sufficient
mass to cause a perturbation of over 200 km\,s$^{-1}$ to the tidal stream,
and VIRGOHI 21 passes to its west, not between it and NGC 4262 as might
be expected if it had pulled the stream westward.  The radial velocity
of `C' means that it must be moving past the stream at a velocity
(relative to the putative undisturbed velocity of the stream)
of at least 350 km\,s$^{-1}$.  At this speed, it would not have stayed close
to the stream long enough to have severely disturbed it.
It would also be expected
that if `C' were involved in the interaction then there would be gas
falling onto it, but this does not appear to be the case as there is no gas
seen between VIRGOHI 21 and C.  The H\,{\sc i} in `C' is unresolved, implying
that it is confined to the area of the optical galaxy; it shows no signs
of infalling gas streams or any other disturbance.  Galaxy `C' is in our a 
minds a possible, but highly unlikely participant in this interaction.

\item Recently Bekki et al. (2005) have published numerical simulations on 
the basis of which they argue that ``VIRGOHI 21 \ldots\ is likely to be tidal 
debris rather than a `dark galaxy'\,''. They 
model high-speed, hyperbolic interactions within cluster potentials (at 350 
and 700 kpc from the cluster center) over 2 Gyr in order 
to draw out very long tidal tails where
the H\,{\sc i} and stars have become separated so that the average surface
brightness within a 100 kpc$^2$ box is $> 30 B$ mag arcsec$^{-2}$ ($L_B <
6.9\times 10^8 L_\odot$), assuming $M_\star/L_B = 4$.
The Bekki et al. stellar density can be converted to $I$-band if we assume
a value for $M_\star/L_I$.  Estimates of this vary (e.g. Portinari et al. 2004;
Vallejo et al. 2002; McGaugh \& de Blok 1997) but are generally close to 1 and 
always less than 2.  We therefore adopt a very conservative (in the sense of
producing the lowest $I$-band surface-brightness) value of $M_\star/L_I = 2$.  
This gives a value of $B-I$ 
for the surface-brightness estimates given in Bekki et al. of 0.75.  For
the two double-tailed models presented (M1 and M2) this gives average
$I$-band surface-brightnesses of 32.7 and 33.0 mag arcsec$^{-2}$, while for
the two single-tailed models presented (M3 and M4) this gives average 
$I$-band surface-brightnesses of 29.5 and 29.9 mag arcsec$^{-2}$.  The 
two-tailed models are inconsistent with our H\,{\sc i} data and the H\,{\sc i}
data of Phookun et al., which clearly show a single tail from NGC 4254, while
the single-tailed models give considerably higher surface-brightnesses (1.2 --
1.6 mags, or 6 -- 8 $\sigma$) than the upper limit from our HST observations.
In addition to this, Bekki et al. find that the velocity fields produced in 
the dark clouds by their models
do not resemble rotation, concluding that ``Velocity fields are the key 
observational tools to help us determine whether gas clouds are (unbound) tidal
debris or are self-gravitating systems embedded within a massive dark matter
halo.''  From both the surface-density of stars expected in a single-tail 
interaction and from the rotation curve, it appears that high-speed, 
hyperbolic interactions within a cluster potential like those proposed by 
Bekki et al. are unable to reproduce our new observations of VIRGOHI 21.

\item NGC 4262, the spiral to the North-East, does not appear to be involved. 
This is not surprising because it is in totally the wrong place to give 
rise to a tidal feature like that observed (NGC 4262, NGC 4254 and VIRGOHI 21 
form a well separted triangle on the sky, see also Vollmer et al. 2005). Also,
the radial velocity between NGC 4254 and NGC 4262 is far
too large (900 km\,s$^{-1}$) to generate tidal features such as bridges and
tails.  Furthermore, with such a large  radial 
velocity difference between NGC 4254 and NGC 4262, any 
interaction must be at high speed and, as seen above from the
simulations of Bekki et al., such interactions are not capable of reproducing 
our observations as even the most isolated gas clouds formed in single-tail 
systems have surface-brightnesses 1.5 - 2 magnitudes brighter than the limit 
set by our HST observations.  NGC 4262 is a S0 galaxy with an 
H\,{\sc i} ring that is inclined with respect to the optical disk (Krumm et al.
1985).  Krumm et al. present higher-resolution H\,{\sc i} observations than
our data and they say that if the H\,{\sc i} has come from an interaction, 
``the narrowness of the ring and the undisturbed appearance of the optical
disk suggest that the donor was a gas-rich dwarf galaxy or intergalactic
H\,{\sc i} cloud, rather than a massive spiral''. Thus we find it very 
difficult to see how NGC 4262 can have given rise to the features we have now 
observed.

\item The components of VIRGOHI 21 are connected both spatially and in
velocity, making it exceedingly improbable that they could be chance
superpositions of smaller hydrogen clouds, while the bridge to NGC 4254 is not 
explained by this hypothesis.

\item Phookun et al. (1993) suggested that the distortion of NGC 4254 
could be due to infalling gas-clouds.  Could VIRGOHI 21 be a tail, similar to 
those seen in \object{UGC 10214} (Briggs et al. 2001) or 
\object{NGC 4038}\object[NGC 4039]{/9} (Gordon et al. 2001)?  In
NGC 4038/9, in particular, there is what appears to be a tidal dwarf  near 
a strong concentration of H\,{\sc i} at the end of the southern tail, around 
60 kpc from the center of the system (for a distance of 13.8 
Mpc; Saviane et al. 2004).  However, the major problem with this hypothesis
 is that NGC 4254 apparently shows
no sign of having undergone a recent merger.  It would be expected that
a merger violent enough to have thrown out an arm over 100 kpc in length
would leave some imprint on the optical disk, but this is not the case here.
Phookun et al. found that ``The galaxy is undergoing an interaction that is
not so strong or violent as to disrupt the disk.''  While this fits with
the hypothesis of VIRGOHI 21 as a dark galaxy, playing a role similar to 
that of the companions of \object{NGC 4027} or \object{NGC 4654} in 
exciting the $m=1$ mode, it does not fit the merger scenario.

\item Oosterloo \& van Gorkom (2005) argue that another H\,{\sc i} 
cloud in the Virgo Cluster, \object{VIRGOHI 4} (Davies et al. 
2004), is caused by ram-pressure stripping from \object{NGC 4388} 
due to an interaction with the hot-gas halo of the \object{M86} sub-group and 
suggest a similar origin could
be possible for VIRGOHI 21.  This would explain the bridge without the need
to invoke a second galaxy, either interacting or merging.  However, we cannot 
see how this can
give the steep, reversed velocity gradient seen in VIRGOHI 21, nor does 
it explain the distortion to the optical disk of NGC 4254. 

\item Single-armed  spirals are normally the result of interactions with 
close-by massive companions (Iye et al. 1982; Phookun et al. 1993);
the lack of any visible companion to NGC 4254 has thus led to a number of
observations and dynamical models.  As mentioned above, recent numerical 
models 
by Vollmer et al. (2005) suggest that NGC 4254 ``had a close and 
rapid encounter
with a $10^{11} M_\odot$ galaxy $\sim 250$ Myr ago.  The tidal interaction 
caused the spiral structure\ldots''.  Phookun et al. (1993), in a VLA 
study of the galaxy, find a trail of gas leading away from it (their Figure 5)
which can be identified from its direction and velocity-gradient as the
NGC 4254 end of the
`bridge' between the two objects.  This supports our  inference that VIRGOHI 21
is the aforesaid $\sim 10^{11} M_\odot$ mass which caused the peculiarities in 
NGC 4254 - particularly so given the lack of any other candidate.

\end{enumerate} 

If our hypothesis is correct that this is a dark, gravitationally bound, 
edge-on rotating disk
then its properties are as presented in Table \ref{properties}.  Judging from 
visible disk galaxies, whose
masses continue to rise beyond their H\,{\sc i} edges (e.g. Salucci \& Persic
1997), the 
full size and mass of such a disk could easily reach the $\sim 10^{11} M_\odot$
required by the Vollmer et al. (2005) simulations.
The very low surface-brightness limits give an upper limit to the luminosity of
the disk of $1.9 \times 10^5 L_\odot$, giving a lower limit for $M_{dyn}/L_B$ 
of at least $10^6 M_\odot/L_\odot$, whereas normal galaxies have values
$< 50$.
%\clearpage
\begin{table}
\caption{Properties of the putative dark disk}
\label{properties}
\begin{tabular}{ll}
Diameter $2R$, [from $+14^\circ 46^\prime$ to $49^\prime$] & 14 kpc\\
Circular Velocity $V_c$, [$(2100 - 1900)/2$] & 100 km\,s$^{-1}$\\
Spin Period $P$, [$2\pi R/V_c$] & $4\times 10^8$ years\\
Total Mass $M_T$, [$RV_c^2/G$] & $2\times 10^{10} M_\odot$\\
Face-on Mass-density [$M_T/\pi R^2$] & $2\times 10^{-2}$ g cm$^{-2}$\\
Hydrogen Mass $M_{HI}$, [$F_{HI} = 0.6$ Jy\,km\,s$^{-1}$]& $3 \times 10^7
M_\odot$\\
Face-on gas density $N_{HI}$, [$M_{HI}/\pi m_{H}R^2$] & $3\times 10^{19}$
cm$^{-2}$\\
Luminosity limit $L_{B}$, [$\mu_B > 32.8$ mag arcsec$^{-2}$]&$<1.9\times 
10^5 L_\odot$\\
Total Mass to Blue Light Ratio $M_{dyn}/L_B$& $> 10^6 M_\odot/L_\odot$\\
\multicolumn{2}{l}{[assuming a disk 0.5 arcmin $\times$ 3 arcmin]}
\end{tabular}
\end{table}
%\clearpage

%\clearpage
\begin{table}
\caption{H\,{\sc i} fluxes and masses of the detected components}
\label{components}
\begin{tabular}{llll}
&$F_{HI}$ (Jy km\,s$^{-1}$)&$M_{HI}$ ($M_\odot$)\\
NGC 4254&91.5&$5.5\times 10^{9}$\\
NGC 4262&7.7&$4.6\times 10^{8}$\\
Object C&0.3&$2\times 10^{7}$\\
`Rotating Disc'&0.6&$3\times 10^{7}$\\
($14^\circ46^\prime<\delta<14^\circ49^\prime$)\\
Low col. density bridge&1.3&$8\times 10^7$\\
($\delta<14^\circ41^\prime$)\\
High col. density bridge&0.8&$5\times 10^{7}$\\
($14^\circ41^\prime<\delta<14^\circ46^\prime$)
\end{tabular}
\end{table}
%\clearpage

It is clear from our new observations that VIRGOHI 21 is intimately linked to 
NGC 4254. The important question is whether it is the cause of the HI bridge 
or the effect of some previous encounter. We believe the weight of the evidence
presented above lies on the side of a dark galaxy.  Dark halos have, after
all, been predicted by galaxy evolution simulations such as Jimenez et al. 
(1997), Verde et al. (2002) and Davies et al. (2006).  That none have been 
confirmed so far is most probably due to instrumental limitations. VIRGOHI 21 
itself was missed in the normal HIPASS-sensitivity observations of this area, 
and only picked up with ten times longer integrations in the  
VIRGOHI survey. In any case single-dish angular resolution makes it all too 
easy to confuse a dark cloud with a bright object clustered with it (Davies et 
al. 2006).  The recent optical identification program for HIPASS 
sources (Doyle et al. 2005) can rule out only `isolated', i.e. 
unclustered, dark galaxies.

If we accept the total mass given in Table 1 then an oddity about VIRGOHI 21 is
its shortage of baryons
compared to visible galaxies. Visible disk galaxies typically have,
in the form of stars and gas, $\sim 10$ percent of their dynamical mass
in baryons, a figure consistent with cosmological models which predict
$\sim 7$ times as much Dark Matter as baryonic in the Universe at large
(Salucci \& Persic 1997).  Had VIRGOHI 21 ten percent of its 
mass in baryons it would contain
$\sim 2\times 10^9$ solar masses which would, if it were once in the form
of H\,{\sc i} spread evenly across the existing dynamical disk, yield a
column density $\sim 10^{21}$ atoms cm$^{-2}$ -- sufficiently above
the ``star-formation threshold'' of $\sim 10^{20}$ cm$^{-2}$ (Martin \&
Kennicutt 2001),
or above the Toomre criterion for gravitational instability in a rotating 
disk (Toomre 1964), to form stars (Jimenez et al. 1997;
Verde et al. 2002).  Since it apparently contains no stars it must have a 
minimum of ten times less baryons - a phenomenon often 
apparent in dwarf galaxies (Mateo 1998).  Thus
90 percent of the baryons one might have expected to find in a typical galaxy
appear to be missing and so it is not difficult to see how a galaxy like this 
may remain dark.
%\newpage

%\ \\

\section{Summary}
In this paper we have presented new Westerbork high resolution 21-cm 
observations of a HI source (VIRGOHI 21) that we have previously observed at 
Jodrell Bank and Arecibo.
These observations clearly show that VIRGOHI 21 has played a part in some form 
of interaction with the bright spiral galaxy NGC 4254. Whether VIRGOHI 21 is 
the result of some previous interaction with a third object or it is the 
interacting object itself is still open to conjecture. We have argued in this 
paper that the weight of evidence now resides on the side of the latter rather 
than the former conjecture. This is primarily because the new observations 
show a long drawn out tidal feature which has an abrupt velocity change and 
we cannot find a candidate galaxy to have caused it. Tidal interactions between
optically bright galaxies often draw out stars as well as gas. We have also 
used HST to search for stars associated with VIRGOHI 21 and find none down to 
very faint surface brightness levels. We look forward to new observations and 
numerical models that will add further weight to either side of this debate.

\acknowledgements
We thank the Netherlands Foundation for Radio Astronomy for use of the
WSRT and Rafaella Morganti and Tom Oosterloo for assistance in reducing the
WSRT data, the National Astronomy and Ionosphere Center for use of the 
Arecibo Telescope, the UK Particle Physics and Astronomy Research Council for 
financial support, and the Australia Telescope National Facility for 
inspiration and technical support.  We would like to thank the following for
useful comments: Virginia Kilborn, Erwin de Blok, Martin Zwaan and Greg 
Bothun.  We also thank the anonymous referee for comments that helped improve
this paper.  The Digitized 
Sky Survey was produced at the Space Telescope Science Institute under U.S. 
Government grant NAG W-2166. The images of these surveys are based on 
photographic data obtained using the Oschin Schmidt Telescope on Palomar 
Mountain and the UK Schmidt Telescope. The plates were processed into the 
present compressed digital form with the permission of these institutions.

\end{document}